\documentclass[mathleft]{an}
\usepackage{graphicx}
\usepackage{subfigure}
\usepackage{times}
\begin{document}

\Pagespan{789}{}
\Yearpublication{2006}%
\Yearsubmission{2005}%
\Month{11}%
\Volume{999}%
\Issue{88}%

\title{Reinvestigation of the electron fraction and \\electron Fermi energy of neutron star}

\author{Z.-F. Gao\inst{1,2}\fnmsep\thanks{Corresponding author:
  \email{zhifugao@xao.ac.cn}\newline}\and H. Shan\inst{1} \and W. Wang\inst{3}\and N. Wang\inst{1}\fnmsep\thanks{Corresponding author:
  \email{na.wang@xao.ac.cn}\newline}}
\titlerunning{Electron Fermi energy}
\authorrunning{Gao et al.}
\institute{Xinjiang Astronomical Observatory, CAS,150, Science 1-Street, Urumqi, Xinjiang, 830011, China
\and Key Laboratory of Radio Astronomy, Chinese Academy of Sciences, West Beijing Road, Nanjing, 210008, China
\and School of Physics and Technology, Wuhan University, Wuhan, Hubei, 430072, China}
\received{5 June 2017}
\accepted{30 August 2017}
\publonline{later}
\abstract{In this work, we reinvestigate the electron fraction $Y_{e}$ and electron Fermi energy $E_{\rm F}(e)$ of neutron stars, based on our previous work, in which we firstly deduced a special solution to $E_{\rm F}(e)$, and then obtained several useful analytical formulae for $Y_{\rm e}$ and matter density $\rho$ within classical models and the relativistic mean field\,(RMF) theory using numerically fitting. The advantages of this work include the following aspects:(1) The linear functions are substituted for the nonlinear exponential functions used in the previous work. This method may be more simple, and closer to realistic equation of state\,(EoS) of a neutron star\,(NS),
because there are linear or quasi-linear relationships between number fractions of leptons and matter density, which can be seen by solving NS EoS; (2)we introduce a dimensionless variable $\varrho$\,($\varrho=\rho/\rho_0$, $\rho_{0}$ is the standard saturated nuclear density), which greatly reduces the scope of the fitting coefficients;(3)we present numerical errors including absolute and relative deviations between the data and fit. By numerically simulating, we have obtained several analytical formulae for $Y_{\rm e}$ and $\rho$ for both APR98 and RMF models. Combining these analytical formulae with the special solution, we can calculate the value of $E_{\rm F}(e)$ for any given matter density. Since $Y_e$ and $E_{\rm F}(e)$ are important in assessing cooling rate of a NS and the possibility of kaon/pion condensation in the NS interior, this study could be useful in the future study on the thermal evolution of a NS.}
\maketitle
\section{Introduction}
Neutron star\,(NS) constitutes one of the best astro-physical
laboratories for studying dense matter physics.  An equation of
state\,(EoS) of matter under exotic conditions is a prerequisite
for studies of the structure and evolution of compact stars. As
an extremely interesting and important physical parameter in NS
EoS, the electron fraction, $Y_{\rm e}$, influences on the
weak-interactions processes, e.g., modified Urca reactions,
$\alpha-$decay, electron capture as well as the absorption of
neutrinos and anti-neutrinos, and etc\,(see Yakovlev et al. 2001; Dong et al. 2013, 2016; Gao et al. 2011a, 2011b, Liu \& Liu 2017a, 2017b; Liu et al. 2017a, 2017b; Sun et al. 2016; Cheng et al. 2015, 2017a, 2017b). The electron fraction $Y_{\rm e}$ is defined as $Y_{\rm e}=n_{\rm e}/n_{\rm B}$, and varies with matter density, where $n_{\rm e}$ and $n_{\rm B}$ are the electron
number density and baryon number density, respectively. For a given NS matter density, how to exactly determine the values of $Y_{e}$ has long been a very challenging task for both the nuclear physics and astrophysics community, due to some uncertainties and artificial assumptions. Currently, our knowledge of $Y_{\rm e}$ mainly comes
from model-dependent EoS of a NS\,(e.g., Baym, Pethick \& Sutherland 1971; Douchin \& Haensel 2000, 2001; Lattimer \& Prakash 2007; Kaminker et al. 2014; Gomes et al. 2014, 2017; Graber et al. 2015, 2017).

The other important parameter is the electron Fermi energy,
$E_{\rm F}(e)$, which denotes the highest energy of electron gas.
As we know, for degenerate and relativistic electrons in $\beta-$
equilibrium, the distribution function $f(E_{\rm e})$ obeys Fermi-Dirac
statistics, and the electron chemical potential $\mu_{\rm e}$ at zero-
temperature is called its Fermi energy.  In the weak-magnetic field
limit, $B\ll B_{\rm cr}$ ($B_{\rm cr}$ is the electron critical
magnetic field), the isoenergetic surface of degenerate and relativistic
electrons is a spherical surface in the momentum space. The Fermi energy
of electrons inside a NS can exert directly impact not only on the
weak-interactions processes, but also on the electron degeneracy pressure counteracting gravity collapse of the star. These impacts will in turn
change intrinsic EoS\,(e.g., Gao et al. 2015; Liu 2016; Zhu et al. 2016), internal structure and heat evolution, and even
influence the whole properties of the star.
Thus, more attention has been paid to the two parameters above due to
their importance.

Our previous work of Li et al.(2016)(``Li16'' in short) is devoted to consideration of the electron Fermi energy and an associated  value of electron number fraction as functions of matter density $\rho$ within different layers of neutron star's crust and core of a common
NS with $B\ll B_{\rm cr}$\,(Li et al. 2016; Gao et al. 2017).  Since the main purpose of this paper is to reinvestigate $Y_e$ and $E_{\rm F}(e)$ inside a NS by revising Li16, it is necessary to briefly review Li16. Please see below for details.

(i)Based on the basic definition of the Fermi energy of degenerate and relativistic electrons, we deduced a special solution to the electron
Fermi energy,
\begin{equation}
E_{\rm F}(e)=60\times(\frac{\rho}{\rho_{0}})^{1/3}(\frac{Y_{e}}{0.005647})^{1/3}~~~~({\rm MeV}).
\label{1}
\end{equation}
which is suitable for relativistic electron matter region ($\rho \geq 10^7$~g cm$^{-3}$) in a common NS. Here $\rho_{0}=2.8\times10^{14}$~g~cm$^{-3}$ is the standard nuclear density. 

(ii) According to generally accepted and reliable EoSs of a NS, we obtained several useful analytical formulae for $Y_{e}$ and matter density $\rho$ within classical matter models and the work of Dutra et al. (2014)(Type-2) in relativistic mean field\,(RMF) theory.

(iii) Since $E_{\rm F}(e)$ and $Y_{\rm e}$ are smooth and continuous functions of $\rho$, we plot the diagrams of $E_{\rm F}(e)$ vs. $\rho$ and $Y_{e}$ vs. $\rho$ using the fitting formula of $E_{\rm F}(e)$, $Y_{e}$ and $\rho$ in four matter regions (the outer crust, inner crust, outer core, and inner core) and some boundary conditions.

(iv) When describing the mean-field Lagrangian, density, we adopted the TMA parameter set, which aims at a consistent description of all the nuclei with one parameter
set and is remarkably consistent with the updated astrophysical observations of NSs. Due to the importance of the
density dependence of symmetry energy, $J$, in nuclear astrophysics, a brief discussion on $J$ and its slop was presented.

(v) Compared with previous studies on the electron Fermi energy in other models, the methods of calculating $E_{\rm F}(e)$ in Li16 are more simple and convenient. Since Urca reactions are expected in the center of a massive star due to high-value $E_{\rm F}(e)$ and $Y_{\rm e}$, the work of Li16 could be useful in the future studies on the NS thermal evolution.

It must be said that our endeavors in Li16 are indeed of practical use
for exploring efficiently many properties of NSs, especially if the fits
are in a compact and widely applicable form. Recently, after a
careful examination, we found some inadequacies of Li16, which should be improved significantly. See below for details.

(i)The electron fractions in Li16 have in fact in all cases expressed as second or third order polynomials in the natural
exponential function of $e^{\rm Log_{10}\rho}$, which leads to an extremely wide coefficient range. Such a wide range of coefficients indeed limits their applicability. It can be implemented by solving a linear system instead of the more
involved non-linear optimization problem needed for expressions
like Eq.(9) of Potekhin et al.\,(2013).

(ii)No mention to the errors in the fit
parameters is presented in Li16.  Although our fit quality is indeed so good that all significant digits in Li16 are basically correct and the fitted errors can be neglected.

Based on the analysis above, we will reinvestigate the fraction
and Fermi energy of relativistic electrons inside a
common NS.  We will perform numerical simulations in APR98 model in second 2, and in RMF models in section 3. The main conclusions are given in section 4.
\section{Numerical Simulations in the APR98 model}
 Employing Argonne v18 two-nucleon\,(Av18) interaction and variational chain summation methods, Akmal,~Pandharipande,~\&~Ravenhall(1998) (``APR98'') investigated the properties of dense nucleon matter and the structure of NSs, and provided an excellent fit to all of the nucleon-nucleon scattering data in the Nijmegen data base (Stoks et al. 1993). In APR98, the authors not only considered the non-relativistic
calculations with Av18 and Av18+UIX\,(Urbana IX three-nucleon interaction) models for nuclear forces, but also
described the relativistic boost interaction model (denoted as $\delta v$) with and without three-nucleon interaction (UIX$^{*}$). The difference between Av18$+\delta v$ and Av18$+\delta v$+UIX$^{*}$ models lies in that whether the effect of three-nucleon interaction\,(TNI) is considered. Here we will choose Av18$+\delta v$ and Av18$+\delta v$+UIX$^{*}$ as two representative models in the following simulations
because these two models can be regarded as more realistic models.
\begin{figure}[bt]
\centering
\includegraphics[angle=0,scale=.88]{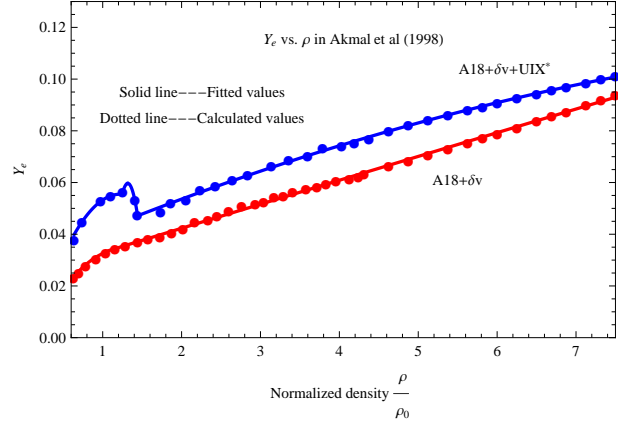}
\caption{Comparisons of the data and fits for $Y_{e}$  as a function
of $\rho$ for for $A18+\delta v$ and $Av18+\delta v+UIX^{*}$ models in APR 1998.}
\label{fig1}
\end{figure}

According to the APR98, the effective interactions have same form
\begin{eqnarray}
H&=&\left[\hbar^2/(2m) + \left(p_3 + (1-Y_p)p_5\right)\rho e^{-p_4 \rho}\right] \tau_n+  \nonumber \\
&&\left(\hbar^2/(2m)+(p_3 + Y_p p_5)\rho e^{-p_4 \rho}\right)\tau_p+ \nonumber \\
 && g(\rho, Y_p=0.5)\left(1-(1 - 2 Y_p)^2\right) + \nonumber \\
&&  g(\rho, Y_p=0) (1- 2 Y_p)^2 ,~~~~~ 
\label{2}
\end{eqnarray}
where $\rho= \rho_n + \rho_p$ at zero temperature, and
\begin{equation}
\tau_p=\frac{3}{5}(3\pi^2\rho)^{\frac{2}{3}}Y_p^{\frac{5}{3}},
\tau_n=\frac{3}{5}(3\pi^2\rho)^{\frac{2}{3}}(1-Y_p)^{\frac{5}{3}}. 
\label{3}
\end{equation}
The parameters defining the $\tau$-dependent
terms are the same for the two models, and are given in APR98.
For Av18+$\delta v$ and three-nucleon interaction models at the low-density phase, the parametrization $g_L(Y_p)$ is expressed as
 \begin{eqnarray}
&&g_L(\rho, Y_p=0.5)=  \nonumber \\
&&-\rho^2(p_1+p_2\rho+p_6\rho^2 + (p_{10}
    +p_{11}\rho){e}^{-p_9^2\rho^2}),~\nonumber \\
&&g_L(\rho, Y_p=0)=\nonumber \\
&&-\rho^2(p_{12}/\rho+ p_7+p_8 \rho+p_{13}\,{e}^{-p_9^2\rho^2}), \label{4}
 \end{eqnarray}
while at the high-density phase, the parameter $g_H(Y_p)$ is written as
\begin{eqnarray}
&&g_H(Y_p=0.5)= g_L(Y_p=0.5)- \nonumber\\
&&\rho^2\left(p_{17}(\rho-p_{19})+p_{21}(\rho-p_{19})^2\right){\rm e}^{p_{18}(\rho-p_{19})}, \nonumber \\
&&g_H(Y_p=0)= g_L(Y_p=0)- \nonumber\\
&&\rho^2\left(p_{15}(\rho-p_{20})+p_{14}(\rho-p_{20})^2\right){\rm e}^{p_{16}(\rho-p_{20})}.
\label{5}
\end{eqnarray}
The parameter values of $P_j, i=1, 2, \cdots 21$ are listed in Table 1. Here, the parameters $p_3=89.8$~MeV fm$^{5}$, $p_4=0.457$~fm$^{3}$ and $p_5=-59.0$~MeV fm$^{5}$ are common to these two models.
\begin{table*}[th]
\caption{Partial values of $n_{\rm B}$, $\rho$, $Y_{\rm e}$ and $E_{\rm F}(e)$ for $Av18+\delta v +UIX^{*}$
and $Av18+\delta v$ models. }
\label{tlab-1}
\footnotesize
\begin{tabular}{ccccccccccc}
\hline
Model&$P_{1}$&$P_{2}$&$P_{6}$&$P_{7}$&$P_{8}$&$P_{9}$&$P_{10}$&$P_{11}$
&$P_{12}$&$P_{13}$\\
\hline
Av18+$\delta v$+UIX$^*$&337.2&$-382.0$&$-19.1$&214.6&$-384.0$
&6.4&69.0&$-33.0$&0.35&0\\
A18+$\delta v$&281.0&$-151.1$&$-10.6$&210.1&$-158.0$
&5.88&58.8&$-15.0$&$-0.2$&$-0.9$\\
 \hline
Model&$p_{14}$&$p_{15}$&$p_{16}$&$p_{17}$&$p_{18}$
$p_{19}$&$p_{20}$&$p_{21}$&$--$&$--$\\
\hline
Av18+$\delta v$+UIX$^*$&0 &287.0 &$-1.54$
&175.0&$-1.45$&0.32&0.195&0&$--$&$--$\\
\hline
\end{tabular}\\
\end{table*}

\begin{table*}[th]
\caption{Partial values of $n_{\rm B}$, $\rho$, $Y_{\rm e}$ and $E_{\rm F}(e)$ for $Av18+\delta v +UIX^{*}$
and $Av18+\delta v$ models. }
\label{tlab-1}
\footnotesize
\begin{tabular}{cccccccccc}
\hline
$n_B$ &Matter-density& $Y_{\rm e}$ & $E_{\rm F}(e)$ & $E_{\rm F}^{\dag}(e)$& $n_B$ &Matter-density& $Y_{\rm e}$ & $E_{\rm F}(e)$ &$E_{\rm F}^{\dag}(e)$  \\
(fm$^{-3}$)  &$(\rm g\, cm^{-3})$ & (\%) & (MeV) & (MeV) &(fm$^{-3}$)  &$(\rm g\, cm^{-3})$ & (\%) & (MeV) & (MeV) \\
\hline
 &  $Av18+\delta v$      &  &   &     & &     &  && \\
 \hline
0.10 & $1.661 \times 10^{14}$ &  2.395 &  81.60  & 81.32 &  0.67 & $1.113 \times 10^{15}$ &  6.237 & 211.64  & 211.96    \\
0.17 & $2.178 \times 10^{14}$ &  2.789 &  102.45 & 102.19 &  0.74 & $1.229 \times 10^{15}$ & 6.620  & 223.15  &  223.33   \\
0.23 & $3.819 \times 10^{14}$ &  3.683 &  124.33  & 124.03&  0.82 & $1.362 \times 10^{15}$ &  7.068 & 236.02  &   236.32   \\
0.30 & $4.982 \times 10^{14}$ &  4.165  & 141.52  & 141.02 & 0.90 & $1.495 \times 10^{15}$ &  7.528 & 248.63  &  248.76       \\
0.37 & $6.144 \times 10^{14}$ & 4.590  &  156.76  & 156.76 &  0.96 & $1.594 \times 10^{15}$ &  7.881 &  257.95  &  258.21 \\
0.41 & $6.808 \times 10^{14}$ &  4.820 &  164.88  &164.56 &  1.00 & $1.661 \times 10^{15}$ &  8.121 &  264.11 & 264.43    \\
0.45 & $7.473 \times 10^{14}$ &  5.043 &  172.67  &172.36 &  1.04 & $1.727 \times 10^{15}$ &  8.365 &  270.23 &  270.65     \\
0.49 & $8.137 \times 10^{14}$ &  5.262  &  180.17  &180.10 &  1.07 & $1.777 \times 10^{15}$ &  8.552 &  274.23 & 274.54     \\
0.57 & $9.465 \times 10^{14}$ &  5.695 &  194.55  & 194.67 &   1.14 & $1.893 \times 10^{15}$ &  8.991 &  285.42 &  285.56     \\
0.64 & $1.063 \times 10^{15}$ &  6.073 & 206.59  & 206.65 & 1.20 & $1.993 \times 10^{15}$ & 9.378   & 294.45 & 294.64  \\
\hline
 &  $A18+\delta v+UIX^{*}$      &  &   &     & &      &  && \\
\hline
0.10 & $1.661\times 10^{14}$ &  3.707  &  94.38  & 94.09 & 0.67 & $1.113\times 10^{15}$ &  7.633  &   226.38  & 226.65\\
0.17 & $2.178\times 10^{14}$ &  4.701  &  121.94  &121.38 & 0.74 & $1.229\times 10^{15}$  & 8.001  &   237.71 & 237.93  \\
0.23 & $3.819\times 10^{14}$ &  4.791  &  135.72 & 135.25 & 0.82 & $1.362\times 10^{15}$ &  8.404  &  250.04   & 250.21 \\
0.30 & $4.982\times 10^{14}$ &  5.274   & 153.11  &152.81 & 0.90 & $1.495\times 10^{15}$  &  8.789  &  261.99  & 262.23  \\
0.37 & $6.144\times 10^{14}$ & 5.796   & 169.44   & 169.17 & 0.96 & $1.594\times 10^{15}$  & 9.068   &   270.29  &270.65 \\
0.41 & $6.808\times 10^{14}$ &  6.073  &  178.09   & 177.88 & 1.00 & $1.661\times 10^{15}$  & 9.251   &  275.84  &  275.99 \\
0.45 & $7.473\times 10^{14}$ &  6.385  &  186.33  & 186.01 & 1.04 & $1.727\times 10^{15}$ & 9.429  &  280.26   & 280.54 \\
0.49 & $8.137\times 10^{14}$ &  6.592   &  194.23  & 194.03 & 1.07 & $1.777\times 10^{15}$ & 9.563  &  285.26  & 285.76  \\
0.57 & $9.465\times 10^{14}$ &  7.073 &  209.12 & 209.37 &  1.14 & $1.893\times 10^{15}$ &   9.864  &  294.37   &  294.57\\
0.64 & $1.063\times 10^{15}$ &  7.468   & 221.54  &221.33 &  1.20 & $1.993\times 10^{15}$ & 10.12   &  301.98 & 302.46\\
\hline
\end{tabular}\\
Footnote: $^{\dag}$. Calculated values of $E_{\rm F}(e)$ by combining Eq.(1) with the fitted polynomial expressions of Eqs.(6-7).
\end{table*}
Table 2 lists partial values of $n_{\rm B}$, $\rho$, $Y_{\rm e}$ and $E_{\rm F}(e)$.
In order to reduce the scope of coefficients of fitted polynomial expressions of $Y_{e}$ and $\rho$ for $A18+\delta v$ and $Av18+\delta v+UIX^{*}$ models, let us introduce a dimensionless variable $\varrho$ ($\varrho=\rho/\rho_0$, $\rho_0$ is the standard saturated nuclear density), which is the order of close to unity, and adopt a new form of $\sum_{n=0}^{n_{\rm max}}c_{n}\varrho^{n}$, where $c_{n}$ is the $n-$th order coefficient of the fitted polynomial expression. As to the $Av18+\delta v+UIX^{*}$ model, the original data of $Y_{\rm e}$ are divided into three groups with consideration of variation tendency. By performing 2nd order polynomial fitting, we obtain
\begin{eqnarray}
 Y_{e}= -0.01232+0.1184\,\varrho- 0.0572\,\varrho^2,\nonumber\\
 Y_{e}= -1.4321+2.3246\,\varrho- 0.9085\,\varrho^2,\nonumber\\
 Y_{e}=0.0291+0.0146\varrho-5.68\times10^{-4}\varrho^2,
 \label{6}
\end{eqnarray}
for $\varrho\sim 0.59-1.19$\,($\rho\sim(1.66\times10^{14}-
3.331\times10^{14}$)\,g~cm$^{-3}$), $\varrho\sim1.190-1.37$\,($\rho\sim(3.33\times10^{14}-
3.83\times10^{14})$\,g~cm$^{-3}$), and $\varrho\sim 1.37-7.12$ ($\rho\sim (3.83\times10^{14}-1.99\times 10^{15})$\,g~cm$^{-3}$),
respectively, where $n_{\rm max}=2$ is assumed. When at the density-node of $\varrho=1.19$, the ``jump'' of $Y_{\rm e}$ is $0.0007$, corresponding to a relative variation $|\Delta Y_{\rm e}/Y_{\rm e}|
\sim1.4\%$, while at the point $\varrho=1.366$, the ``jump'' is about
0.0006, corresponding to a relative variation of $1.2\%$. Thus, the
continuities of three functionals in Equation (6) are enforced. The
comparisons between the data and the fit show that the residuals of $Y_{\rm e}$ are typically $10^{-4}$,the relative differences are smaller than $3\%$.
\section{Numerical simulations in the RMF models}
To date, many of relativistic models have drawn attentions in investigating NS EoSs (e.g., Glendenning et al. 1985; Schaffner \&~Mishustin 1996; Zhou et al. 2017; Mu et al. 2017). The most common among them is the RMF theory, which has become standard method to study nuclear matter and finite-nuclei properties. According to RMF models, the strong interaction between baryons is mediated by
the exchange of isoscalar scalar and vector mesons
$\sigma$, $\omega$, isovector vector meson $\rho$.
There are two additional strange mesons namely isoscalar
scalar $\sigma^{*}$ and vector $\phi$ mesons considered
by some authors (e.g., Yang \& Shen 2008; Xu et al. 2012; Zhao 2015, 2016). The total effective Lagrangian is given by
\begin{eqnarray}
L&&=\sum_B\overline{\psi}_B[i\gamma_\mu\partial^\mu-(m_B-g_{\sigma B}\sigma-g_{\sigma^*B\sigma^*})- \nonumber\\
&&g_{\rho B}\gamma_{\mu}{\mathbf{\tau}}\cdot{\mathbf{\rho}^\mu}-g_{\omega B}\gamma_\mu\omega^\mu-g_{\phi B}
\gamma_\mu\phi^\mu]\psi_B+ \frac{1}{2}\nonumber\\
&&(\partial_\mu\sigma\partial^\mu\sigma-m_\sigma^2\sigma^2)+\frac{1}
{2}(\partial_v\sigma^*\partial^v\sigma^*-m^2_{\sigma^*}\sigma^{*2}) \nonumber\\
&&-\frac{1}{4}W^{\mu v}W_{\mu v}-\frac{1}{4}R^{\mu v}R_{\mu v}+\frac{1}{2}m_\rho^2{\mathbf{\rho}}_\mu{\mathbf{\rho}}^\mu-\frac{1}{4}P^{\mu v}  \nonumber\\
&&P_{\mu v}+\frac{1}{2}m_\omega^2\omega_\mu\omega^\mu-\frac{1}{3}a\sigma^{3}-\frac{1}{4}b\sigma^4
+\frac{1}{2}m^2_{\phi}\phi_\mu\phi^\mu \nonumber\\
&&+\frac{1}{4}c_3(\omega_\mu\omega^\mu)^2+\sum_l\overline{\psi}_l[i\gamma_\mu\partial^\mu-m_l]\psi_l.
\label{7}
\end{eqnarray}
where $W_{\mu v}=\partial_\mu\omega_v-\partial_v\omega_\mu$,
$R_{\mu v}=\partial_\mu{\mathbf{\rho}}_v-\partial_v{\mathbf{\rho}}_\mu$
and $P_{\mu v}=\partial_\mu\phi_v-\partial_v\phi_\mu$ denote the
field tensors of $\omega$, $\rho$ and $\phi$ mesons, respectively, and
$\gamma_{u}$ is the Dirac matric. The meson field equations in uniform matter have the following form
\begin{eqnarray}
\sum_B g_{\sigma B}\rho_{SB}=m_\sigma^2\sigma+a\sigma^2+b\sigma^3,~\sum_B g_{\omega B}\rho_B\nonumber\\
=m_\omega^2\omega_0+c_{3}\omega^{3}_{0},~\sum_B g_{\rho B}\rho_{B}I_{3B}=m_{\rho}^2\rho_0,\nonumber\\
 \sum_B g_{\sigma^* B}\rho_{SB}=m_{\sigma^*}^2\sigma^*,\sum_B g_{\phi B}\rho_{B}=m_\phi^2\phi_0,
 \label{8}
\end{eqnarray}
were $J_{B}$ and $I_{3B}$ express the baryon spin and isospin projections, respectively. $m_B^{*}$ is the baryon effective mass
\begin{equation}
m_B^{*}=m_B-g_{\sigma B}\cdot\sigma-g_{\sigma^* B}\cdot\sigma^*,
\label{9}
\end{equation}
At zero temperature the lepton chemical potentials are expressed as
\begin{equation}
\mu _{l}=\sqrt{{k_{F}^{l}}^{2}+m_{l}^{2}}, ~~~~(\rm fm^{-1}).
\label{10}
\end{equation}
The charge neutrality condition is given by
\begin{equation}
\sum_{B}q_{B}\rho_{B}-n_{e}-n_{\mu}=0,
\label{11}
\end{equation}
where $\sum_{B}q_{\rm B}=n_{\rm B}$, and $q_{\rm B}$ is the baryon electric charge.
The coupled equations can be solved self-consistently.

In order to numerically simulate in RMF models, we select three representative parameter-set groups: NL3, TMA and GM1(SU3). The former two include three measons: $\sigma$, $\omega$ and $\rho$, while the latter one includes $\sigma$, $\omega$, $\rho$, $\sigma^{*}$ and $\phi$. These three RMF parameter-set group are successful in describing NS matter to some extent\,(Glendenning  1985; Lalazissis et al. 1995; Toki et al. 1995; Geng et al. 2005).

The saturation properties including mass parameters, meson-nucleon couplings and self-coupling constants of three RMF parameter sets are listed in Table 3.
\begin{table*}[th]
\caption{Saturation properties, meson-nucleon couplings and self-coupling constants of three RMF models. }
\label{tlab-2}
\begin{tabular}{ccccccccccc}
\hline
Model & $\rho_0$ & $E_0$ & $K_0$ & $m^{*}$ & $K^{'}$ & $J$ & $L_{0}$ &
$K^{0}_{sym}$ &$Q^{0}_{sym}$ & $K^{0}_{\tau,V}$ \\
      & fm$^{-3}$ & MeV & MeV & & MeV & MeV& MeV & MeV & MeV & MeV \\
\hline
NL3   & 0.148 &-16.24 & 271.53 &0.60 &-202.91 &37,40 &118.53 & 100.88&
181.31 & -698.85 \\
TMA & 0.147 &-16.02 &318.15&0.635 &572.12 &30.66& 90.14&10.75&-108.74&-367.99 \\
GM1(SU3)&0.153&-16.33&300.50&0.70&215.66&32.52&94.02&17.98&25.01&-478.64\\
\hline
Model&$m_N$ & $m_\sigma$ & $m_\omega$ & $m_\rho$ & $g_{\sigma N}$& $g_{\omega N}$ & $g_{\rho N}$ & $a$  & $b$ &$c_3$  \\
 & MeV & MeV & MeV & MeV  & fm$^{-1}$ & fm$^{-1}$ & fm$^{-1}$ &fm$^{-1}$ & fm$^{-1}$ &fm$^{-1}$ \\
\hline
 NL3 &939.0&508.194&782.50&763.0&10.217&12.868&4.474&-10.431&-28.885&0 \\
 TMA&939.0 & 519.151 & 781.95 & 768.1 & 10.055 & 12.842 & 3.800 & $-0.328$ & 38.862 & 151.590\\
 GM1(SU3)$^{\dag}$&938.0&550&783.0&770.0&4.10&10.26&4.10 &12.28&-8.98&0
 \\
\hline
\end{tabular}\\
Footnote:$^{\dag}$. For the GM1(SU3) parameter set, the meson masses $m_{\sigma*}=975.0$\,MeV, and $m_{\phi*}=1020.0$\,MeV, the meson-
hyperon couplings $g_{\sigma \Lambda}=6.170$\,fm$^{-1}$, $g_{\sigma \Xi}=1020.0$\,fm$^{-1}$,$g_{\sigma* \Lambda}=5.412$\,fm$^{-1}$, and $g_{\sigma* \Lambda}=11.516$\,fm$^{-1}$.
\end{table*}

In Table 3, $\rho_0$ is the saturation density,
$E_0$=$(E/A)_\infty$ is the bulk binding energy/nucleon, $K_0$ is the incompressibility, $m^{*}=M^{*}/M$ is the effective mass ratio,
$K^{'}=-Q_0$ ($Q_0$ is the skewness coefficient), $J$ is the symmetry energy at $\rho=\rho_0$, $L_{0}$ is the slope of the symmetry energy($S$), $K^{0}_{sym}$ is the curvature of $S$, $Q^{0}_{sym}$ is the skewness of $S$ and $K^{0}_{\tau,V}$ is the volume part of the isospin
incompressibility.

Inserting the parameter values of TMA into the standard procedure of RMF models, we calculate the values of the related quantities for TMA parameter set. By fitting, we obtain the analytical representations of $Y_{\rm e}$ and $\rho$ in TMA model
\begin{eqnarray}
Y_ {\rm e}=-0.00316+0.05258\,\varrho-0.00514\,\varrho^{2}, \nonumber \\
 Y_ {\rm e}=0.08235+0.0124\,\varrho-5.04\times10^{-4}\,\varrho^{2},
 \label{12}
\end{eqnarray}
for $\rho\sim(6.92\times10^{11}-9.38\times10^{14})$ and
$(9.38\times10^{14}-2.69\times10^{15})$~g~cm$^{-3}$, respectively.
At the midpoint of $2.98\times10^{14}$\,g~cm$^{-3}$, the ''jump'' of $Y_{\rm e}$ is about $2.8\times10^{-3}$, and its relative variation
$\sim2.5\%$ confirming the continuities of two expressions above. The typical differences between the fit and the data are $10^{-3}-10^{-4}$, and their relative differences are typically $10^{-2}-10^{-3}$. The maximum absolute deviation and relative error are $4.5\times10^{-3}$ and $3\%$, respectively, at the high-density end, due to uncertainty of the EoS. Combining Equation (12) with Equation (1), we can calculate the value of $E_{\rm F}(e)$ in any given matter density for TMA parameter set. The comparisons of the EoS and its analytical expressions are shown in Figure 2(a) and (b).
For the GM1(SU3) parameter set, the relation $g_{\sigma^{*}N}=g_{\rho_{\Lambda}}=0$ is assumed.
Inserting the parameter values of GM1(SU3) into the standard procedure of RMF models, we calculate the values of $n_{\rm B}$, $n_{\rm e}$, $Y_{\rm e}$ and $E_{\rm F}(e)$. By fitting, we obtain the analytical representations of $Y_{\rm e}$ and $\rho$ in GM1(SU3) model
\begin{eqnarray}
 Y_{\rm e}=-0.00298+0.0526\,\varrho-0.00494\,\varrho^{2},\nonumber\\
 Y_{\rm e}=0.07663+0.0138\,\varrho-5.99\times 10^{-4}\,\varrho^{2},
 \label{13}
\end{eqnarray}
for $\rho\sim(6.92\times10^{11}-8.04\times10^{14})$ and
$(8.04\times10^{14}-2.69\times10^{15})$~g~cm$^{-3}$, respectively.
At the midpoint of $8.04\times10^{14}$\,g~cm$^{-3}$, the ''jump'' of $Y_{\rm e}$ is about $4\times10^{-3}$, and its relative variation $\sim3.7\%$, which also ensures the continuities of two expressions above. The typical differences between the fit and the data are $10^{-3}-10^{-4}$, and their relative differences are typically $10^{-3}$.

In the same way, we calculate the values of the related quantities for NL3 parameter set. The analytical expressions of $Y_{\rm e}$ and $\rho$ for NL3 parameter set are given as
\begin{eqnarray}
 Y_{\rm e}=-0.00436+0.0749\,\varrho-0.00851\,\varrho^{2},\nonumber\\
 Y_{\rm e}=0.11556+0.00931\,\varrho-3.52\times 10^{-4}\,\varrho^{2},
  \label{14}
\end{eqnarray}
for $\rho\sim (5.69\times10^{11}-7.42\times10^{14})$ and
$(7.42\times10^{14}-2.71\times10^{15})$~g~cm$^{-3}$, respectively.
At the midpoint of $7.42\times10^{14}$\,g~cm$^{-3}$, the ''jump'' of $Y_{\rm e}$ is about $3.4\times10^{-3}$, and its relative variation $\sim2.5\%$, which also ensures the continuities of two expressions above. The maximum absolute deviation and relative error are $3.5\times10^{-3}$ and $2\%$, respectively, at the high-density end.

Using Equations.(13-14) and Equation (1), we obtain the fitted values of $E_{\rm F}(e)$ for both NL3 and GM1(SU3) parameter sets. Figure 2(c-f) shows comparisons of the EoS of the two parameter sets and their analytical representations. In Figure 2, the data points are rarefied, and the dots and lines are for the data and the fit, respectively. It is also obvious that the analytic results agrees well with the data obtained from these three parameter sets.
\begin{figure}[!htbp]
\centering
 \includegraphics[width=0.45\textwidth]{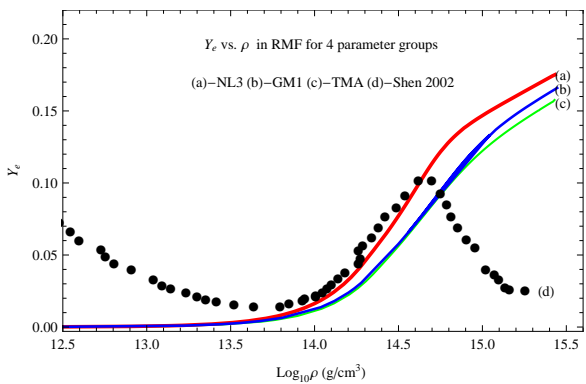}
\caption{Comparing $Y_{e}$ vs. $\rho$ of in RMF for 4 parameter groups.}
\label{fig_2}
\end{figure}

In order compare the three parameter sets, we plot the diagrams of $Y_{\rm e}$ and $\rho$ for RMF models in Fig.3. In the figure, the added red-dot-line is fitted from the work of Shen (2002). The author constructed the EoS in a wide NS density range using RMF theory. At lower densities, the Thomas-Fermi approximation was used to describe the nonuniform matter composed of a lattice of heavy nuclei; while at higher densities, the TM1 parameter set was adopted. Thus, $Y_{\rm e}$ firstly increased, then decreased with $\rho$. However, the inclusion of hyperons softened the EoS considerably at high densities, the maximum stellar mass in Shen (2002) was estimated about 1.6\,$M_{\rm Sun}$,(due to the depression of hyperons on Fermions), which deviates from the observational NS mass\,(Demorest et al. 2010).
\begin{figure*}[th] %
\begin{center}
\begin{tabular}{cc}
\scalebox{0.76}{\includegraphics{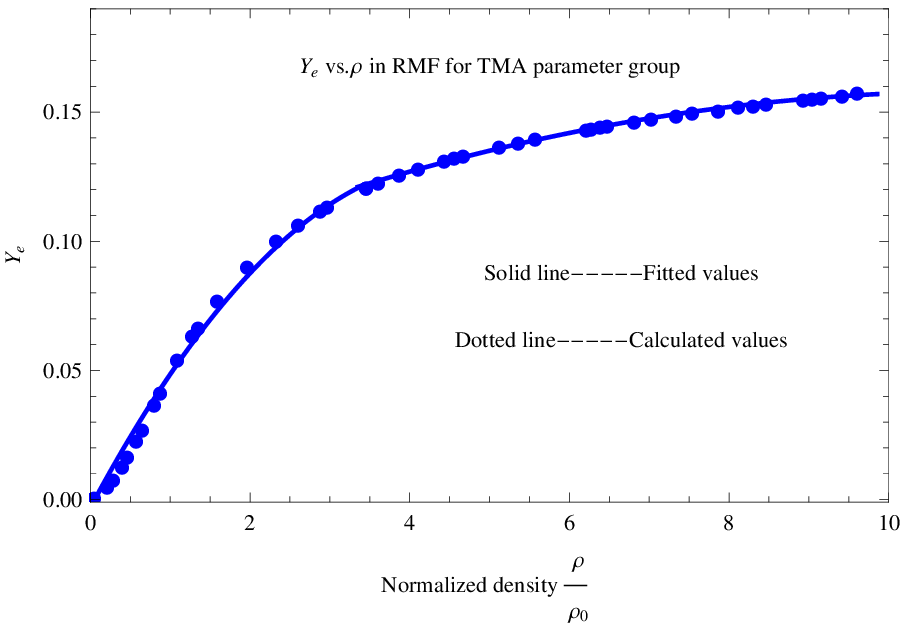}}&\scalebox{0.76}{\includegraphics{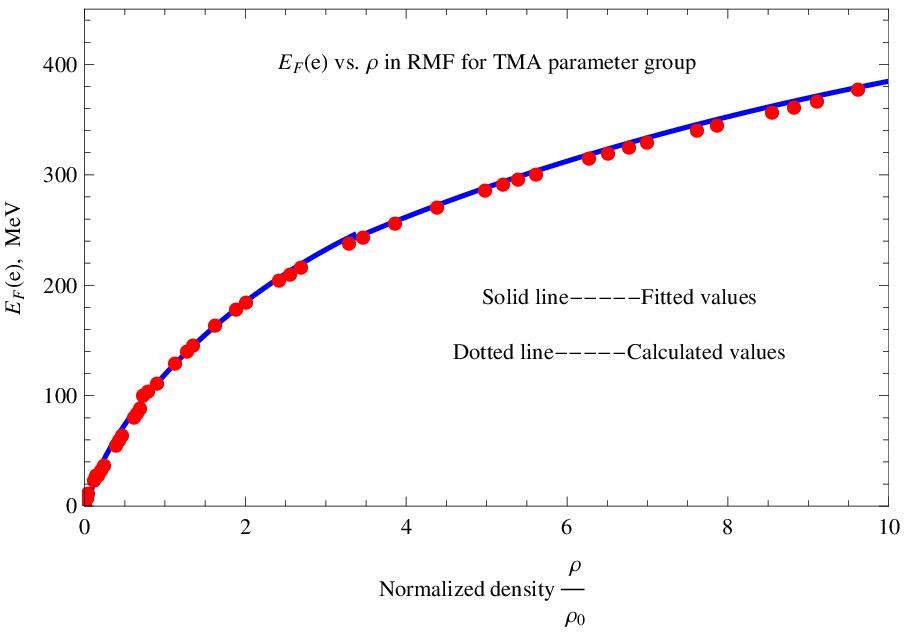}}\\
(a)&(b)\\
\scalebox{0.76}{\includegraphics{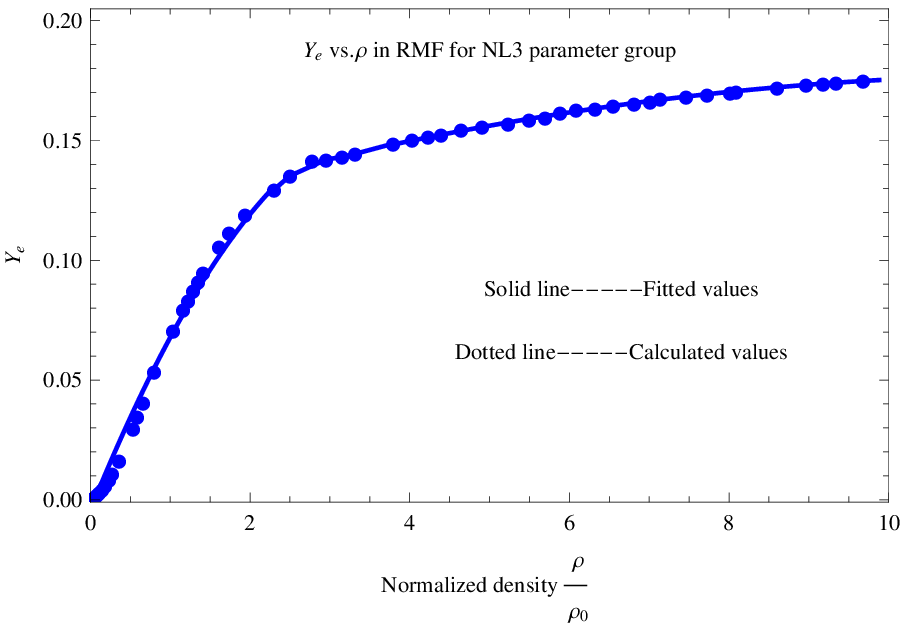}}&\scalebox{0.76}{\includegraphics{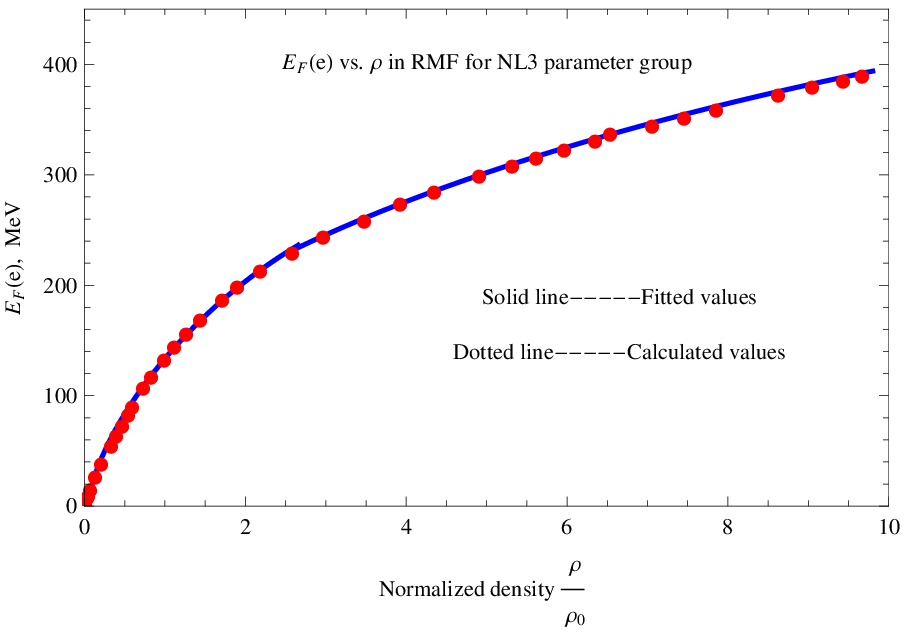}}\\
(c)&(d)\\
\scalebox{0.76}{\includegraphics{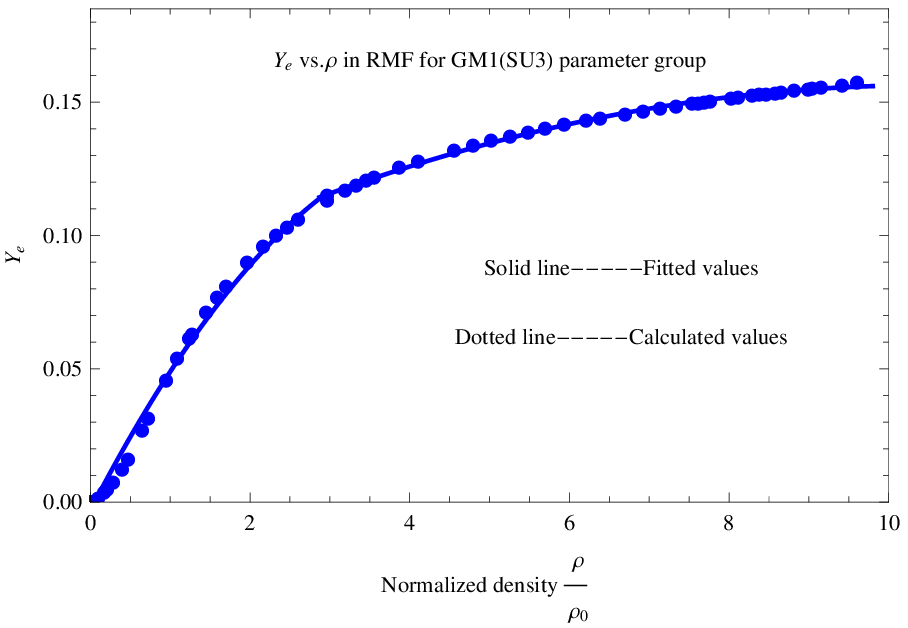}}&\scalebox{0.76}{\includegraphics{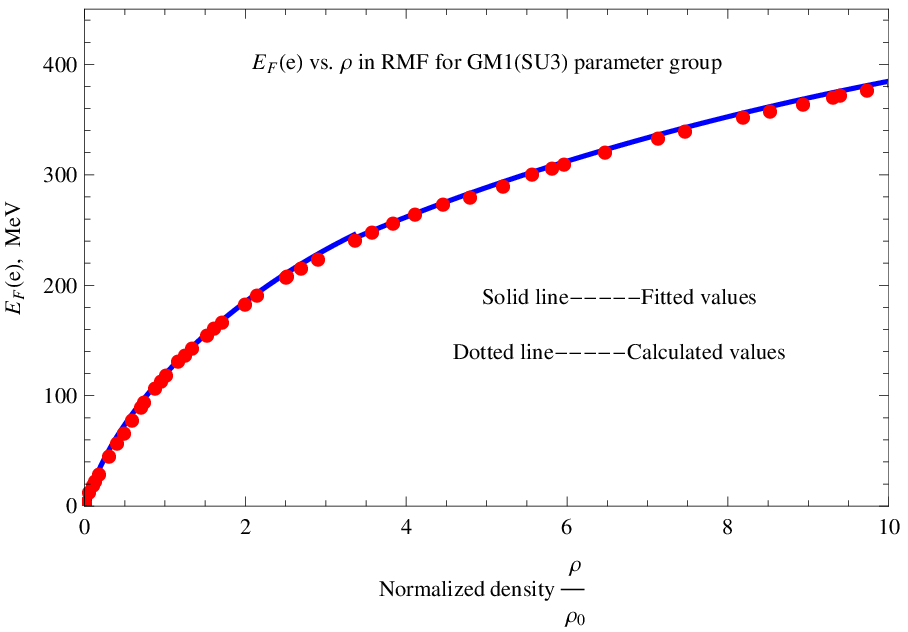}}\\
(e)&(f)\\
\end{tabular}
\end{center}
\caption{Numerically fitting in RMF within TMA, NL3 and GM1(SU3) parameter sets. Left (a), (c), (e), the relations of $Y_e$ and $\rho$. Bottom, (b), (d), (f), the relation of $E_{\rm F}(e)$ and $\rho$. Top, center and bottom are for TMA, NL3 and GM1(SU3) parameter sets, respectively.
 \label{11:fig}}
\end{figure*}
\section{ Discussion and conclusions}
In summary, we have performed numerical simulations firstly in APR98, then in relativistic mean field models, and obtained several analytical representations of $Y_{\rm e}$.  Since $Y_{\rm e}$ and $E_{\rm F}(e)$ are important in assessing cooling rate of a NS and the possibility of kaon/pion condensation in the NS interior, the analytical representations obtained will be very useful in the future study on thermal evolution of a NS and the EoS of NS matter under extreme conditions, though our methods are indeed simple.
\acknowledgements
This work was supported by National Basic Research Program of China grants 973 Programs 2015CB857100, the West Light
Foundation of CAS through grants XBBS-2014-23, XBBS-2014-22 and 2172201302, Chinese National Science Foundation through grants No.11673056,11622326, 11273051,11373006, 11133004 and 11173042, the Strategic Priority Research Program of CAS through XDB23000000 and National Program on Key Research and Development Project through No. 2016YFA0400803.

\end{document}